\documentclass[]{emulateapj} 
\usepackage[colorlinks=true,urlcolor=blue,citecolor=blue,linkcolor=blue]{hyperref}
\usepackage{graphicx}
\usepackage{url}
\usepackage[usenames]{color}
\usepackage{natbib}
\usepackage{amsmath}

\def\CaII{\ion{Ca}{2}}
\def\OI{\ion{O}{1}}
\def\Oline{\OI~135.6$\,$nm line}
\def\MgII{\ion{Mg}{2}}
\def\HI{\ion{H}{1}}
\def\HeI{\ion{He}{1}}
\def\HeI1083{\HeI\ 1083 nm}

\def\MgII{\ion{Mg}{2}}
\def\MgIIk{\ion{Mg}{2}\,k}

\def\hk{{h\&k}}
\def\MgIIhk{\MgII\, \hk}
\def\kthree{\mbox{k$_3$}}    

\def\ktwo{\mbox{k$_2$}}

\def\MgIIkthree{\MgII\,\kthree}
\def\MgIIktwo{\MgII\,\ktwo}
\def\MgIIkline{\MgIIk~line}
\def\MgIIkwing{\MgIIk~wing}
\def\MgIIkprofiles{\MgIIk~profiles}
\def\MgIIkcore{\MgIIk~emission~core}

\def\Cair{\CaII\ 854.2 nm}
\def\CaIIHK{\CaII\,H\&K}
\def\Halpha{\mbox{H$\alpha$}}
\def\Lyalpha{\mbox{Ly$\alpha$}}
\def\Lybeta{\mbox{Ly$\beta$}}

\def\kms{\mbox{km s$^{-1}$}}

\newcommand{\be}{\begin{equation}}
\newcommand{\ee}{\end{equation}}

\newcommand{\bea}{\begin{eqnarray}}
\newcommand{\eea}{\end{eqnarray}}

\begin{document}

\title{What do IRIS observations of \MgIIk\ tell us about the solar plage chromosphere?}
  
  \author{Mats Carlsson$^{1}$}\email{mats.carlsson@astro.uio.no}
   \author{Jorrit Leenaarts$^{2}$}\email{jorrit.leenaarts@astro.su.se}
    \author{Bart De Pontieu$^{3,1}$}\email{bdp@lmsal.com}

\affil{$^1$ Institute of
  Theoretical Astrophysics, University of Oslo, P.O. Box 1029
  Blindern, NO--0315 Oslo, Norway} 
\affil{$^2$ Institute for Solar Physics, Department of Astronomy,
  Stockholm University,
AlbaNova University Centre, SE-106 91 Stockholm, Sweden}
\affil{$^3$Lockheed Martin Solar \& Astrophysics Lab,
         Org.\ A021S, Bldg.\ 252, 3251 Hanover Street,
         Palo Alto CA~94304, USA}

\date{Received 2015 June 30; accepted 2015 July 30}

\begin{abstract}
We analyze observations from the Interface Region Imaging Spectrograph
(IRIS) of the \MgIIkline, the \MgII\ UV subordinate lines, and the
\Oline\ to better understand the solar plage chromosphere. We also
make comparisons with observations from the Swedish 1-m Solar
Telescope (SST) of the H$\alpha$ line, the \ion{Ca}{2} 8542 line and
{\it Solar Dynamics Observatory}/Atmospheric Imaging Assembly observations of the coronal 19.3$\,$nm line. To understand the
observed \MgII\ profiles, we compare these observations to the
results of numerical experiments.
The single-peaked or flat-topped \MgIIkprofiles\ found in plage imply
a transition region at a high column mass and a hot and dense
chromosphere of about 6500 K. This scenario is supported by the
observed large-scale correlation between moss brightness and
filled-in profiles with very little or absent self-reversal.
The large wing width found in plage
also implies a hot and dense chromosphere with a steep chromospheric
temperature rise. The absence of emission in the \MgII\ subordinate
lines constrain the chromospheric temperature and the height of the
temperature rise while the width of the \Oline\ sets a limit to the
non-thermal velocities to around 7\,\kms.
\end{abstract}

  
\section{Introduction}                          \label{sec:introduction}



Plages are regions of the solar atmosphere with strong unipolar magnetic field. The Interface Region Imaging
Spectrograph mission 
\citep[IRIS;][]{2014SoPh..289.2733D}, 
provides new diagnostics for the properties of plage in the form of the  \MgIIhk\ lines, the \MgII\ subordinate 
UV lines, which are sensitive to heating in the low chromosphere
 \citep{2015ApJ...806...14P};
and the \ion{O}{1} line at 135.56$\,$nm that samples the chromosphere but forms under optically thin conditions
\citep{Lin+Carlsson2015}.
In this Letter we investigate what the IRIS observations, together with observations from the 
Swedish 1-m Solar Telescope
\citep[SST;][]{2003SPIE.4853..341S}
and the Atmospheric Imaging Assembly 
\citep[AIA;][]{2012SoPh..275...17L}
on board the {\it Solar Dynamics Observatory (SDO)}
can tell 
us about the heating of the chromosphere and the neighboring regions (transition region (TR) and corona). 

Understanding the physical mechanisms driving the formation of 
\MgIIhk\ is of great interest, not only for understanding the
heating in the chromosphere, but also because these lines are used extensively as a
proxy for the Sun's cyclical variability over the entire UV spectral
domain
 \citep[i.e., the ``\MgII\ index'';][]{Heath1986}. 
%

\section{Observations} \label{sec:observations}

We used four data sets of active regions (including sunspots, pores, and
plage regions) obtained with IRIS,
one of
which (2014 June 11 at 07:36 UT) was coordinated with \Halpha\ and \Cair\ spectral line scans
from the CRisp Imaging SPectrapolarimeter 
\citep[CRISP;][]{2006A&A...447.1111S,2008ApJ...689L..69S}
mounted on the SST.
%
%
All IRIS raster scans were dense
($0\farcs35$ steps) with a spatial sampling of $0\farcs17$ along the
slit (unless otherwise noted). The 2014 June 11 data set focused on
NOAA active region (AR) 12080 and consists of
20 large dense rasters with 96 raster steps resulting in a
field of view of 33\farcs5$\times$182\arcsec\ centered at $(x,y)=(573\farcs5,-199\farcs3)$. The exposure time per raster step was
4$\,$s. The raster cadence was 516$\,$s.

The other IRIS data sets are similar large dense rasters with 400
raster steps and a resulting field of view of
140\arcsec$\times$182\arcsec. They have different exposure times, raster durations, and
pointings of, respectively, 
30s, 3.5 hr, $(x,y)=(-113\farcs2,-234\farcs0)$ (NOAA AR 12104)  for the 2014 July 4, 11:40 UT data set; 
8s, 1 hr, $(x,y)=(-108\arcsec,106\arcsec)$ (NOAA AR 12139) for the 2014 August 16, 21:17 UT data set; 
and 30s, 3.5 hr, $(x,y)=(13\farcs3,-243\farcs4)$ (NOAA AR 12187) for the 2014 October 18, 06:45 UT data set. 
The last data set included spatial binning by two along the
slit to improve the signal-to-noise of the weak \OI\ 135.6$\,$nm
line. All data sets were co-aligned with images in the 17.1 and
19.3 nm AIA passbands.
\begin{figure*}
  \includegraphics[width=\textwidth]{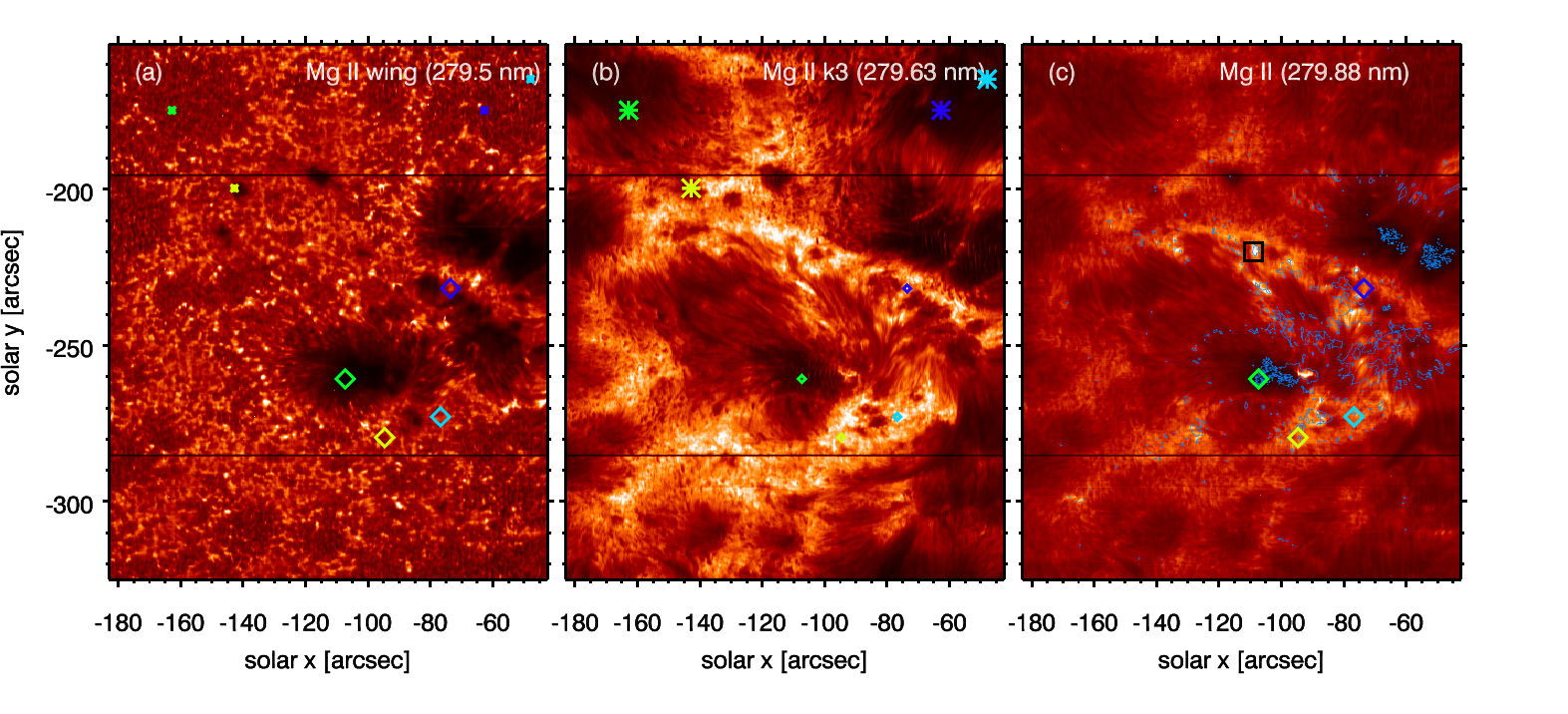}
   \includegraphics[width=\textwidth]{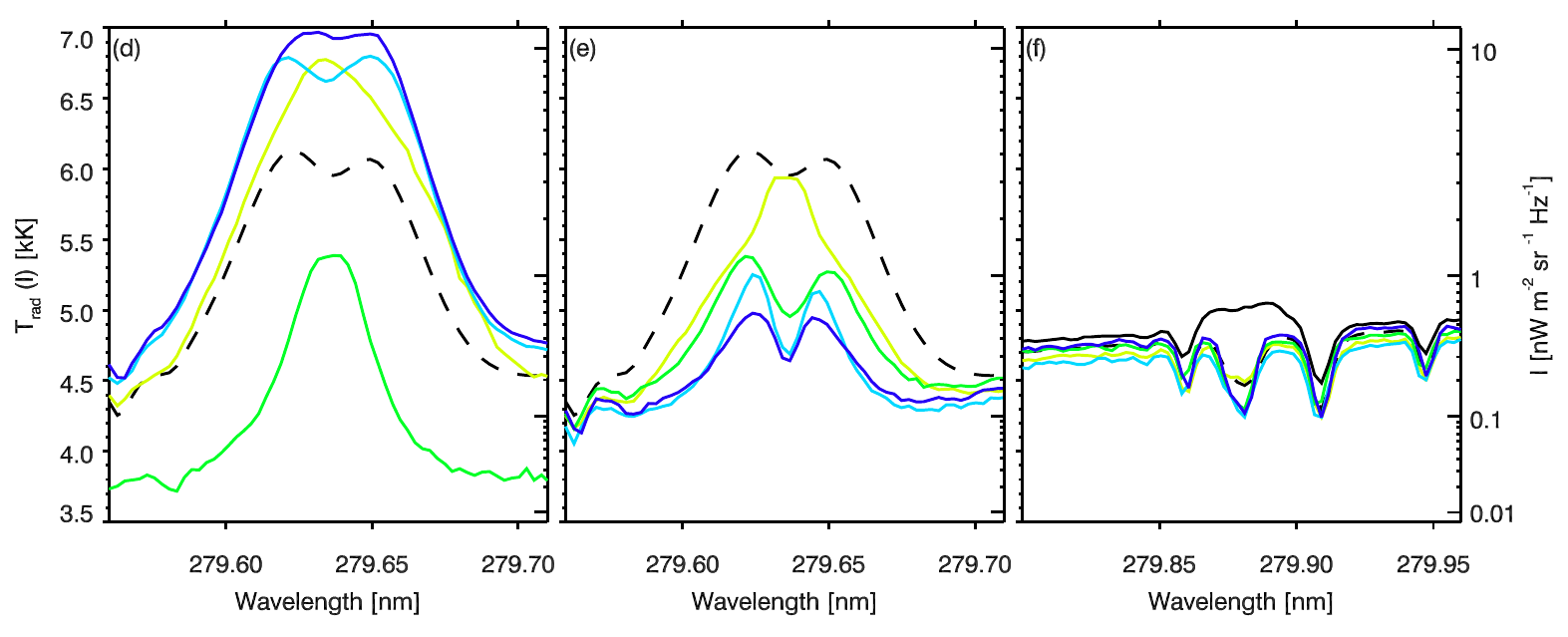}
    \caption{IRIS observations of AR 12104 in the wing of the \MgIIkline\
      (a), at the \MgIIk\  core (\kthree) (b) and in the subordinate
      \MgII\ blend at 279.88 nm (c). Typical plage \MgIIkprofiles\
      are shown in (d) for the
      diamond locations in (a); with non-plage \MgIIkprofiles\ in
      (e) for the star locations in (b). 
      Sample plage profiles of the subordinate \MgII\ 
      279.88 nm blend are in (f) for the same 
      diamond locations as in (a). 
      Full field of view average profile dashed in panels (d)-(f).
      Light-blue contours in (c) show where the subordinate blend
      is in emission, the full black line in (f) is an example profile at the black
      square location in (c). 
      \label{fig:images}} 
 \end{figure*}

\section{Observed Plage Properties} \label{sec:plage}

\subsection{Typical  \MgIIk\ Profiles}

The \MgIIkline\ shows a wide range of profiles throughout AR 12104 (Fig.~\ref{fig:images}). While the
average profile of the field of view shows the typical central reversal
(although significantly reduced compared to the quiet Sun), 
this average hides a wide range of types
of profiles. Quiet-Sun regions in the vicinity of a plage (which is bright in \MgII\ wing;
Fig.~\ref{fig:images}(a)) show deep central reversals with typically
fainter \ktwo\ peaks \citep[for a definition of spectral features,
see][]{2013ApJ...772...89L} 
and mostly narrower profiles (blue, cyan, and green profiles/locations
in Fig.~\ref{fig:images}(b),(e)). Sunspots (green in
Fig.~\ref{fig:images}(a),(d)) and pores (yellow in
Fig.~\ref{fig:images}(b),(e)) are fainter, much narrower than the average
AR profile (dashed lines in Fig.~\ref{fig:images}(d)--(f)) and typically show a
single-peak profile. Profiles in plage are very different from quiet-Sun 
and sunspot/pore profiles: they are brighter, wider, and
typically either single-peak (yellow, Fig.~\ref{fig:images}(a),(d)),
flat-topped (blue, Fig.~\ref{fig:images}(a),(d)) or with a very small central
reversal (cyan, Fig.~\ref{fig:images}(a),(d)).

Figure~\ref{fig:images}(c) shows that
\MgII\ 279.88 nm is indeed sensitive to the chromosphere with
many morphological features similar to those seen in \MgIIkthree\
(b). While some locations show
this subordinate blend in emission (light blue contours in (c); black
square/profile in (c)/(f)), they are typically in sunspots,
explosive events such as bombs \citep{2014Sci...346C.315P} or
locations in the canopy that surround the plage. The subordinate blend
almost never goes into emission in the plage regions. This
appears to provide strong constraints on the chromospheric temperature
profile (see \S~\ref{sec:modeling}).

\begin{figure*}
  \includegraphics[width=\textwidth]{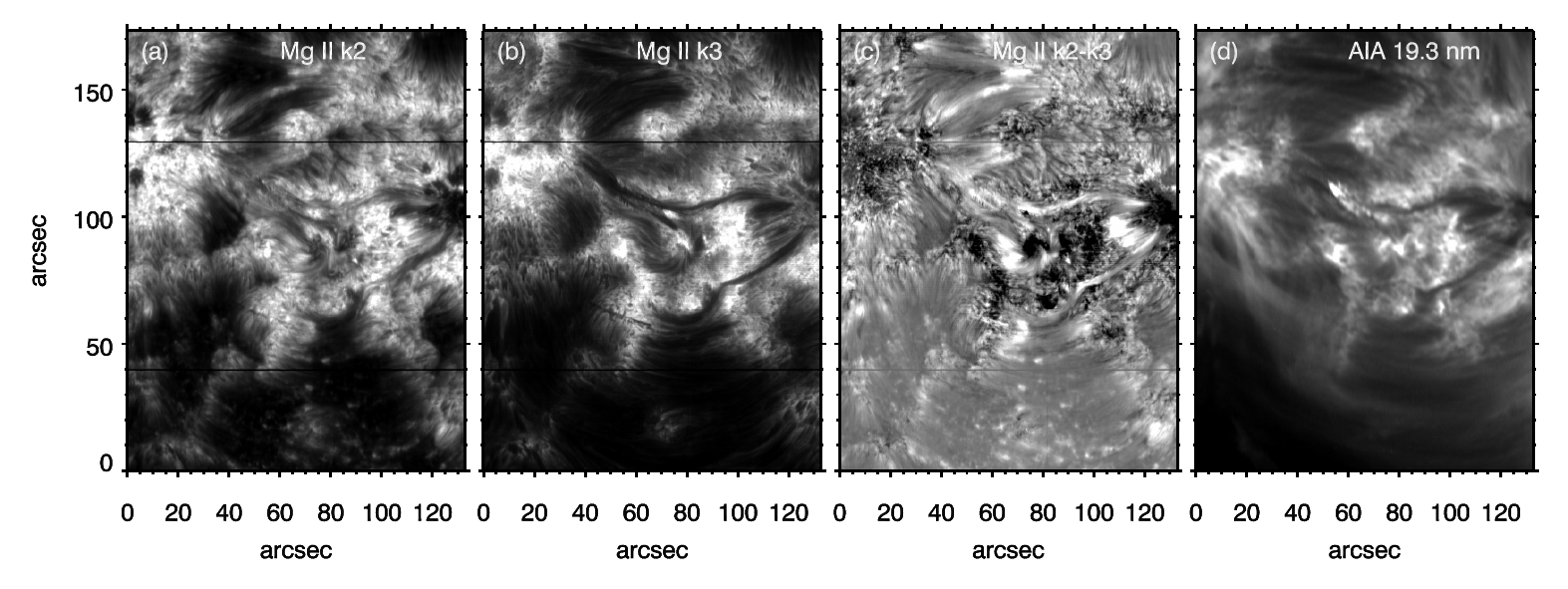}
   \includegraphics[width=\textwidth]{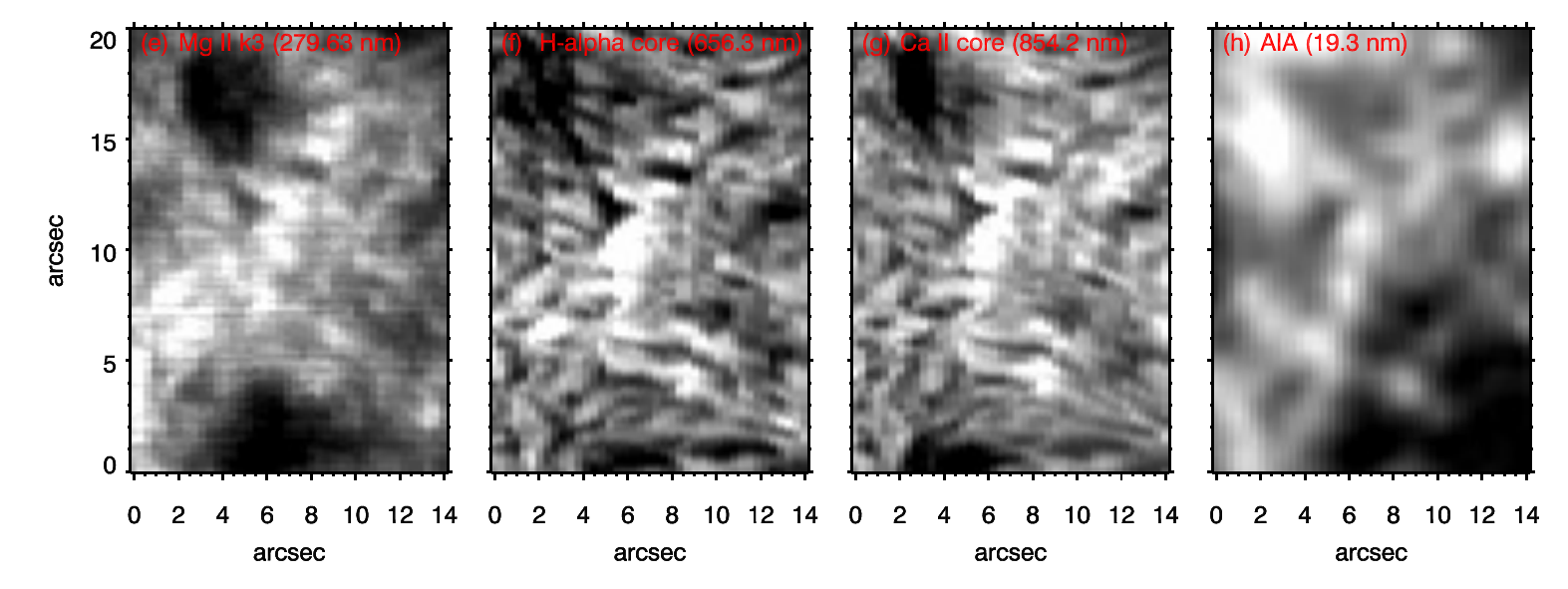}
    \caption{Correlations and comparisons between various \MgIIk\ profile parameters
      and SST and {\it SDO}/AIA observables. Top row: \MgIIktwo\ (a) and
      \MgIIkthree\ (b) of AR 12139, the difference
      between \ktwo\ and \kthree\ (c) and AIA 19.3 nm (d). Bottom row: \MgIIkthree\ intensity of a small plage
      region of AR 12080 (IRIS, e),
      \Halpha\ line core (SST, (f)), Ca II
      854.2 line core (SST, (g)) and AIA 19.3 nm ({\it SDO}, (h)). This figure
      is accompanied by two online animations. \label{fig:correlation}} 
 \end{figure*}
While plage profiles without a central reversal are common, not all
plage regions show such profiles. 
On large, active-region-size, spatial scales, most plage regions are
bright in \ktwo (defined as average brightness in
two fixed wavelength ranges to the blue/red of the line-center
wavelength, as derived from the average profile;
Fig~\ref{fig:correlation}(a)). 
On the same scales, the \kthree\ map (brightness at the
wavelength of line center of the average profile; Fig~\ref{fig:correlation}(b)) shows more variability with some
bright \ktwo\ regions associated with fainter \kthree\ intensity, and
others showing very bright \kthree. This is illustrated by
panel (c): for single-peak profiles \kthree\
intensity is larger than \ktwo, so regions with many single-peak
profiles appear dark in panel (c). In summary, most plage regions are bright in
\ktwo, but only a subset show more single-peak profiles. 

\subsection{Correlations with the TR and corona on large scales}

Comparison with the {\it SDO}/AIA
19.3 passband (dominated by \ion{Fe}{12}, $\log{T_\mathrm{max}} \approx 6.2$; Fig.~\ref{fig:correlation}(d)) shows that
such single-peak regions have a spatial correlation, on large spatial scales, with regions of
enhanced moss emission. This AIA passband contains both bright coronal
loops, 
 dark absorbing features, 
and bright, low-lying moss (e.g.,
(100\arcsec, 95\arcsec)). Moss regions are the upper TR footpoints
of high pressure coronal loops that are so hot that their TR
emission is formed at 1 MK, so that the {\it SDO}/AIA 19.3$\,$nm emission
originates from low-lying plasma that occurs at the same heights as chromospheric
jets or dynamic fibrils 
\citep{1999ApJ...519L..97B,1999ApJ...520L.135F, 1999SoPh..190..419D} 
that absorb some of the EUV emission and give the mottled
appearance typical for moss regions (see, e.g., bottom row of Fig.~\ref{fig:correlation}). 

The brightness of moss is thought to be a good proxy for and linearly
related to the pressure in the overlying hot coronal loops
\citep{2000ApJ...537..471M}. We thus find that the brightest moss
regions, i.e., locations with high coronal pressure, and thus
TR at higher column mass, typically show more
single-peak profiles. 

\subsection{Correlations with chromosphere and TR on small scales}

We also find correlations on the smallest observable spatial scales (bottom row of Fig.~\ref{fig:correlation}). We find that on sub-arcsecond scales
the brightness in a small mossy plage region
of \MgIIkthree, \Halpha\ line center, and \Cair\ are
well correlated. This correspondence is in part due to the small dark
features visible in all three lines that are associated with dynamic fibrils
\citep{2006ApJ...647L..73H,2007ApJ...655..624D}. This indicates that
these diagnostics are all sensitive
to upper chromospheric conditions, including velocity fields. The
\Cair\ brightness and \Halpha\ width have been proposed as a proxy for chromospheric
temperatures in more quiescent conditions
\citep{2009A&A...503..577C}. This relationship seems doubtful in plage:
(1) the \Halpha\ width is not well correlated with
\Cair\ brightness (see the animation of
Fig.~\ref{fig:correlation}); (2) \Cair\ is well correlated
with \MgIIkthree\ brightness, which as shown below appears to be
more sensitive to the TR rather than chromospheric conditions.

At small spatial scales, there is also a good, though not perfect, correspondence with AIA 19.3 nm
moss emission, especially with dark moss features. This is perhaps not
surprising as these features have been previously associated with
bound-free absorption from dynamic fibrils
\citep{1999SoPh..190..419D}. On the other hand, many of the bright
moss locations have an equivalent bright region in \MgIIkthree,
\Halpha\ line center, and \Cair\ line center. These findings expand on the previously found
relationship between upper TR moss emission and \Halpha\ line center
\citep{2003ApJ...590..502D} or Lyman-$\alpha$ emission
\citep{2001ApJ...563..374V}.

\subsection{Properties of Mg II k in Plage}

\begin{figure*}
  \includegraphics[width=\textwidth]{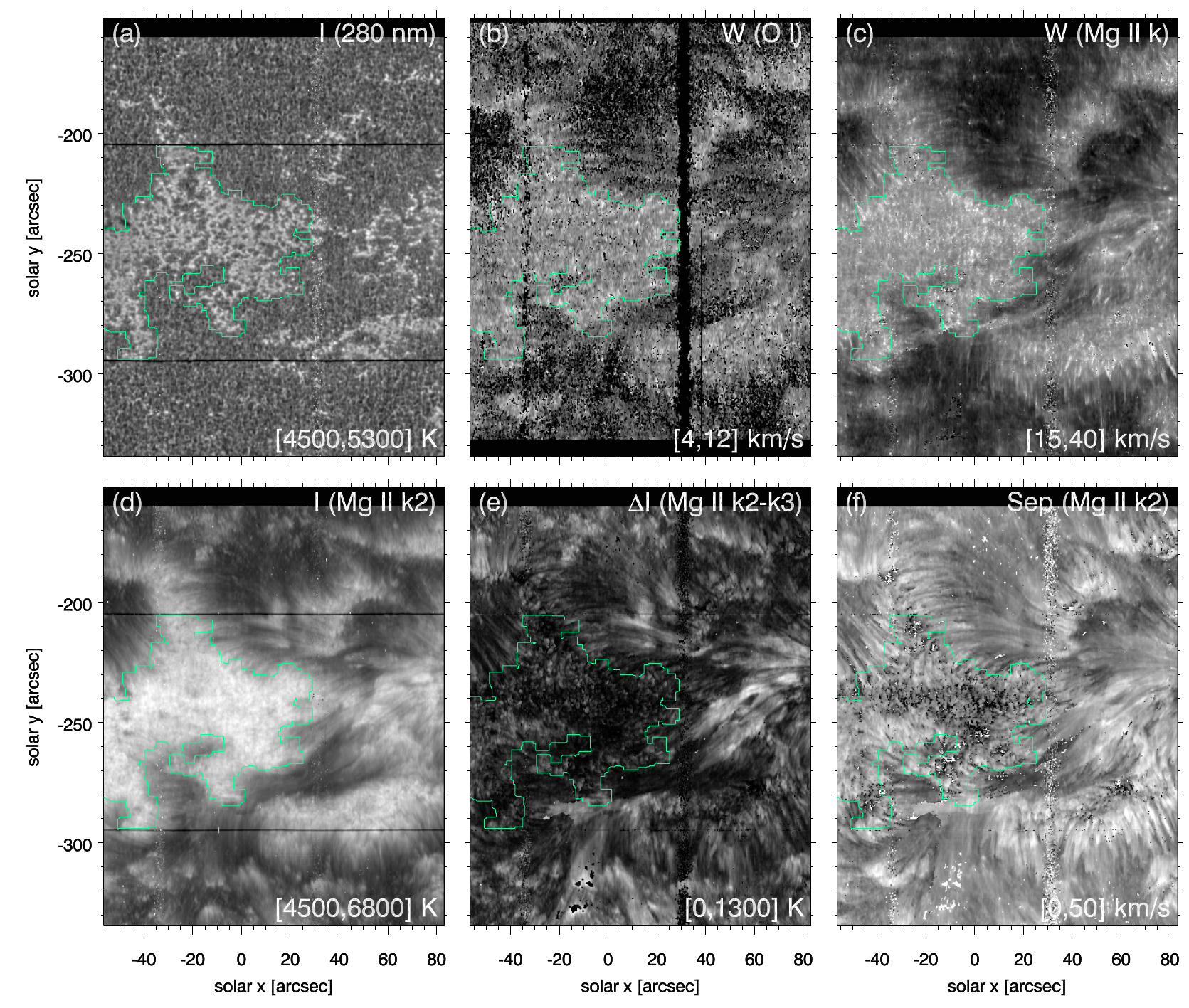}
   \includegraphics[width=\textwidth]{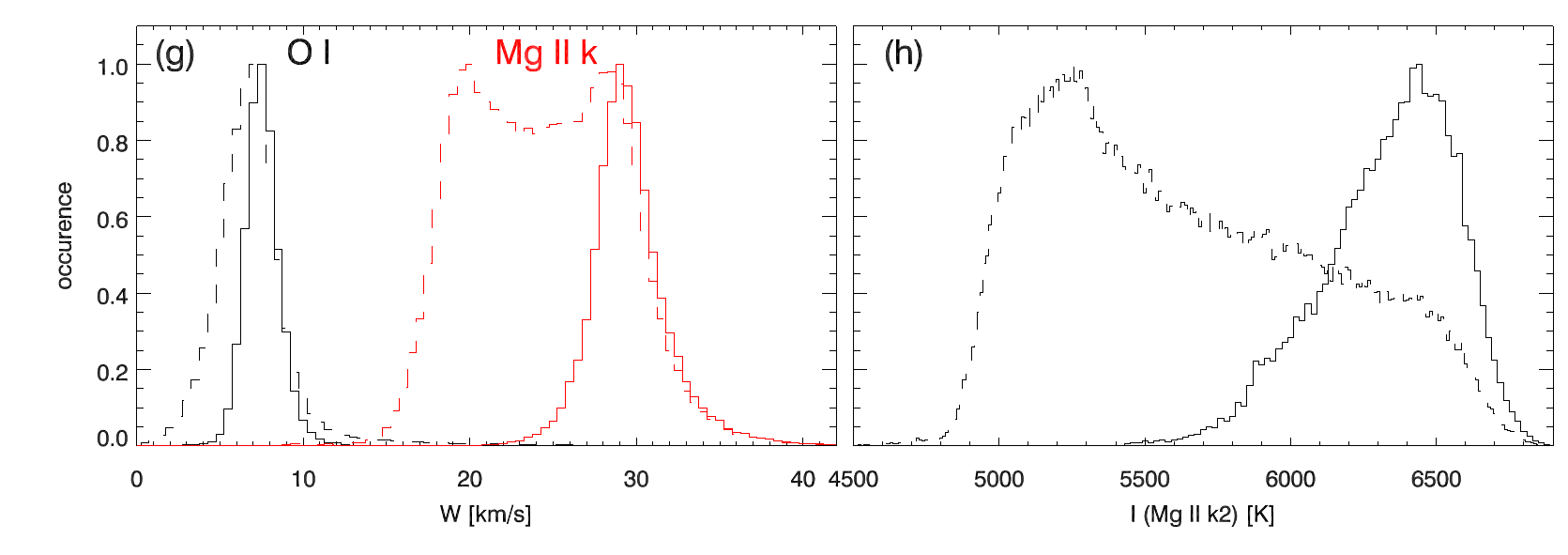}
    \caption{AR 12187 observables: 
    (a) Intensity at 280 nm (photospheric context);
    (b) 1/e width of a single-Gaussian fit to the \Oline.
    (c) 1/e width of single-Gaussian fit to the outer wings of \MgIIk\ (core between \ktwo\ peaks excluded),
    (d) mean intensity of \MgII~\ktwo\ peaks (or maximum intensity for single peaks), 
    (e) difference between peak intensity of the \MgIIkline\ and \kthree\ central depression,
    (f) peak separation between \MgIIktwo\ peaks, 
    (g) histogram of 1/e widths of \OI\ 135.6$\,$nm  (black) and \MgIIkline\ wing fit (red) for the full field of view
     (dashed) and plage (solid), 
    (h) histogram of the radiation temperature of the \MgIIktwo\ peaks for full field of view (dashed) and plage area (solid).
    All intensities are in radiation temperature. Green contours in panels
    (a)-(f) show the plage region, with 
    ranges for the color table in the lower right corner.
    \label{fig:raster}} 
 \end{figure*}

Figure~\ref{fig:raster} shows various observables of AR 12187. 
Panel (a) shows context by showing the intensity at 280 nm, 
a wavelength between the \MgIIhk\ lines where the intensity is formed in the upper photosphere. 
Magnetic areas are clearly seen as increased intensity, and there is a
large plage area in the left part. We use this intensity to determine
a mask for plage (green contours). 

The width of the \Oline\ (panel (b)) in the plage area is remarkably constant,
around 7.8~\kms\ with a small spread (panel (g)), and is larger than in internetwork areas.
The \Oline\ is optically thin, and the width gives a direct measure of the width of the atomic absorption profile \citep{Lin+Carlsson2015}. 
At a temperature of 7\,kK the thermal 1/e width of the \Oline\ is 2.7\,\kms\ leading to a mean non-thermal width of 7.3\,\kms. 

The wing width of the \MgIIkline\ (panel (c)) is similarly larger in the plage area than in non-plage areas
with a mean 1/e width of 30~\kms\ 
with a small spread (g).
The large wing width of the \MgIIkline\ comes from a large "opacity broadening factor" (see
\citet{Rathore+Carlsson2015} for a discussion). 
The wings outside the \ktwo\ peaks sample the velocity field in a similar region as the \Oline. 
There is a clear correlation between the widths of the two lines but not one-to-one, showing that the \MgIIkwing\ width is not only influenced
by the velocity field in the chromosphere but also by a varying opacity broadening factor.

The radiation temperature of the \MgIIk\ peak intensity is around 6400~K in plage with a smaller spread compared with the full field of view
(panels (d),(h)). 
The plage area shows a mottled appearance with very little correlation with the intensity at 280$\,$nm on small scales. 
There is some correlation with the \MgIIkwing\ width (c).

Panel (e) shows how the central reversal of the \MgIIkline\ is filled in by showing the difference 
in radiation temperature between the peak intensity and the \kthree\ intensity. 
Single-peak profiles thus have a difference of zero. 
Most of the central part of the plage has a very weak central reversal or single-peak profiles. 
These locations also have larger maximum intensity (d) and larger \MgIIkwing\ width (c). 
Stronger self-reversals occur in small patches outside the central part of the plage.
These patches are correlated with asymmetric peaks, larger peak intensity, and larger \ktwo\ peak separation 
(panel (f)) but smaller \MgIIkwing\ width (c). 

The \MgIIktwo\ peak separation (panel (f)) shows a lot more structure than the \MgIIkwing\ width (c).
The separation is large for the asymmetric profiles that have a rather deep central reversal (bright patches in the
non-central part of the plage in panel (e)).

\section{Modeling} \label{sec:modeling}

The \MgIIhk\ line profiles from plage regions show some common features: their emission cores are wider than in other atmospheric regions, they are brighter, and their central depressions
are shallower or absent. 

By trial and error we constructed a model that matches the observed average \MgIIk\ profile reasonably well by simultaneously solving the non-LTE problem for hydrogen, calcium, and magnesium, including charge conservation and enforcing hydrostatic equilibrium using the RH code
\citep{2001ApJ...557..389U}.
For hydrogen and calcium, we used the standard 5-level-plus-continuum \HI\ and \CaII\ models that come with RH; for magnesium, we used the 10-level-plus-continuum \MgII\ atom from
\citet{2013ApJ...772...89L}.
The \Lyalpha, \Lybeta, \CaII\,H\&K and \MgIIhk\ lines were computed including partial redistribution; all other lines were computed assuming complete redistribution. 

\begin{figure*}
  \includegraphics[width=\textwidth]{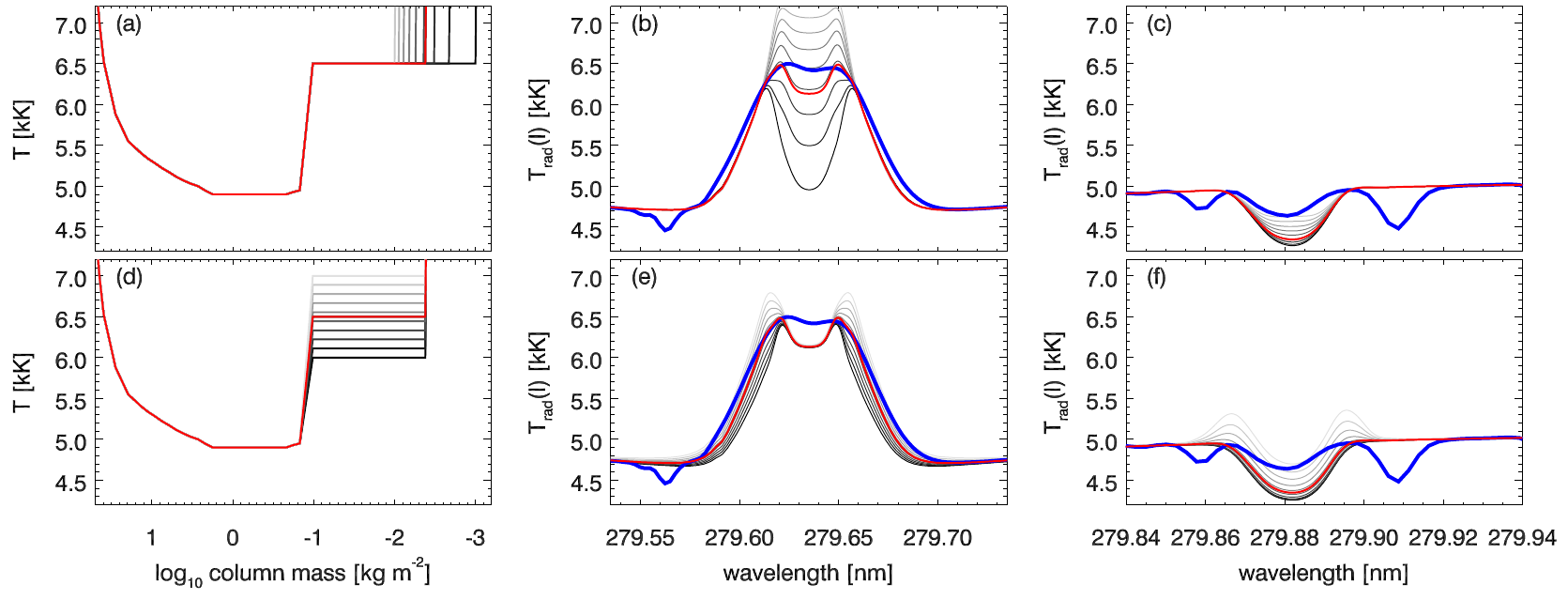}
    \caption{Sensitivity of the emergent \MgIIkline\ to variations in a 1D static plage atmosphere model. Top row: (a) best-fit model atmosphere (red) as function of column mass. Curves in shades of gray show variations of the column mass of the TR. (b) Emergent \MgIIkline-core profile from the best-fit atmosphere (red) and the variations  shown in the panel on the left (corresponding shades of gray). The average plage profile from the IRIS observations is shown in blue. The outer minima are photospheric lines from
different elements than Mg that are not included in the modeling. (c) Same as (b), but now for the two subordinate lines at  279.875 nm and 279.882 nm.
    Bottom row: same as the top row, but now for variations in the temperature of the chromospheric plateau.
   \label{fig:model1}} 
 \end{figure*}

\begin{figure*}
  \includegraphics[width=\textwidth]{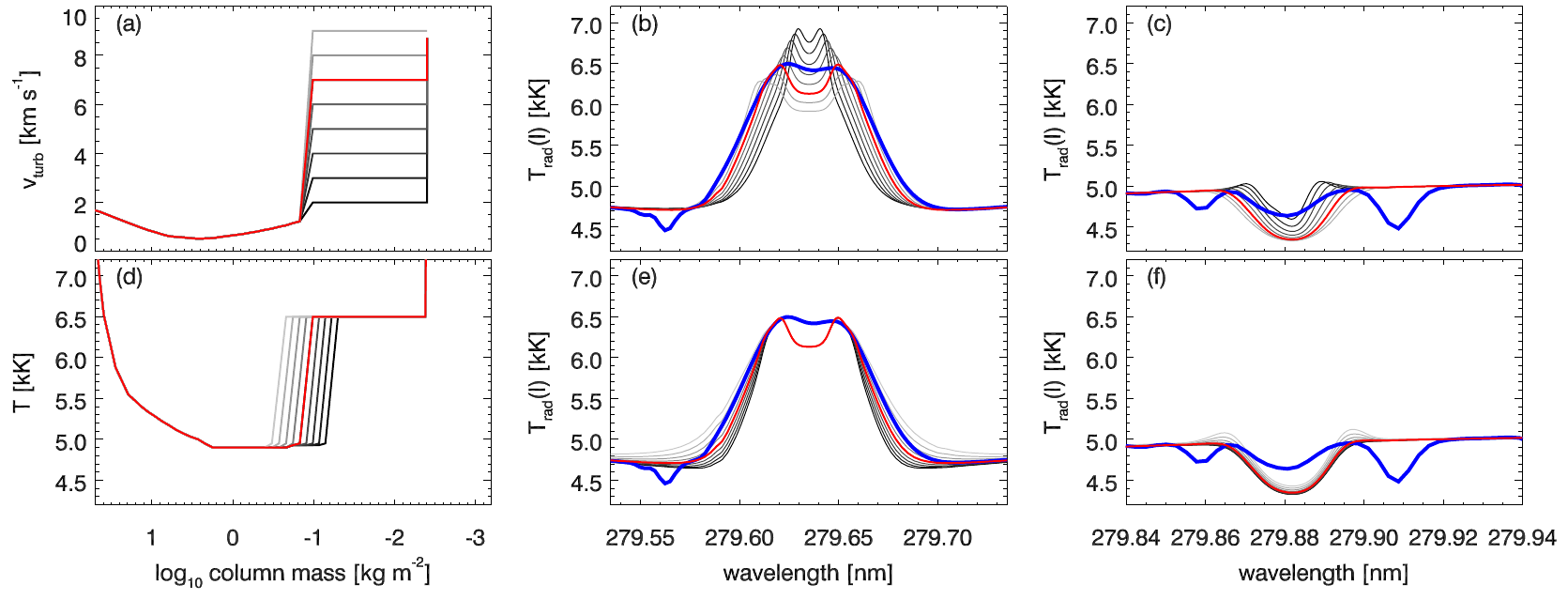}
    \caption{Sensitivity of the emergent lines to variations in a 1D static plage atmosphere model. This figure follows the same format as Figure~\ref{fig:model1}.  Top row: variation in chromospheric microturbulence.  Bottom row: variation in column mass of the chromospheric temperature increase.
        \label{fig:model2}}
 \end{figure*}

The model is constructed from the photosphere and TR of the one-dimensional static FALP model  
\citep{1991ApJ...377..712F}.
Compared to FALP, our model has an extended temperature minimum, a steeper chromospheric temperature rise, a constant chromospheric temperature plateau, and the TR located at a larger column mass. In addition, we replaced the FALP non-constant microturbulence in the temperature plateau with a constant value.
In Figure~\ref{fig:model1} and~\ref{fig:model2} we show the main parameters of our model and compare the synthetic \MgIIkline\-core and the subordinate blend at 279.88~nm with the average observed plage spectrum. 

We do not propose this model as a realistic model of the atmospheric structure of plage; it is a numerical experiment exploring the constraints that the \MgII\ profiles set on the structure of plage chromospheres. We therefore explored the sensitivity of the \MgII\ lines to variations of our model by varying the column mass of the TR (panels (a)--(c) in Figure~\ref{fig:model1}), the temperature of the chromospheric plateau (panels (d)--(f) in Figure~\ref{fig:model1}), the microturbulence in the chromospheric plateau (panels (a)--(c) in Figure~\ref{fig:model2}), and the column mass of the chromospheric temperature rise (panels (d)--(f) in Figure~\ref{fig:model2}).

Varying the location in column mass of the TR has a large effect on \ktwo\ and \kthree: the emission peaks get higher, and the \kthree\ minimum fills up as the TR is moved toward larger column mass, but even the best-fit model still has some central reversal. The subordinate blend gets shallower.

The temperature of the chromospheric plateau has a significant effect on the \MgIIk\ core-width: larger temperatures mean larger widths. For temperatures above 6.5~kK the \ktwo~intensity increases. The subordinate blend reacts strongly to changes in temperature. For temperatures below 6.5~kK, it is in absorption; for higher temperatures, it gets marked emission peaks.

Varying the microturbulence in the chromospheric temperature plateau changes the width of the \MgIIkcore, and the width of the absorption line caused by the UV subordinate transitions.

Changing the location of the chromospheric temperature rise has two effects. First, the deeper in the atmosphere it is located, the wider is the \MgIIkcore. Second, deeply located temperature rises lead to emission in the wings of the subordinate blend.

\section{Discussion and conclusions} \label{sec:discussion}

Our results on the peculiar \MgIIkprofiles\ found in
plage provide an intriguing picture of heating in
the plage chromosphere and hotter layers above.
Our numerical experiments suggest that single-peaked or
flat-topped profiles found in plage arise naturally if
the chromosphere is hot and dense, thus ensuring coupling of the source function to the temperature, and if the column mass of the TR is high so that \ktwo\ and \kthree\ form at the base of the TR. 
The latter occurs naturally at the footpoints of hot,
dense coronal loops in which a strong thermal conductive flux pushes the
TR to high column mass. This scenario is supported by
the observed large-scale correlation between moss brightness (a proxy for
coronal pressure) and filled-in profiles with very little or absent self-reversal.
The correlation is not perfect since it
depends on both chromospheric and coronal conditions. For example, cooler ($\sim1-2$ MK) but dense coronal
loops can also have high coronal pressures (and thus TR at high column
mass), yet they would lack signatures of AIA 19.3 nm ``moss''. In
addition, locations with TR at high column mass may occur even for
a chromosphere that is lower density or cooler (leading to centrally
reversed profiles). Despite these caveats, we often do find a
reasonable correlation on small, arcsecond spatial scales between
\kthree\ brightness and upper TR moss emission.
At these small scales, the \Halpha\ width does not correlate with 
\kthree\ brightness, but the \Cair\ brightness does. This is unlike the quiet-Sun.

We find that, contrary to quiet Sun profiles
\citep{2013ApJ...772...89L,2013ApJ...772...90L}, the \kthree\ and \ktwo\
properties in plage are sensitive to conditions higher up, at the very top of the chromosphere. The \MgIIk\ sensitivity to
\hbox{(mid-)}chromospheric conditions can be found in the width of the \MgIIk\
wing, the (lack of) emission of the subordinate blend at 279.88
nm and the \ktwo\ peak separation. Compared to quiet Sun, the \MgIIkwing\ width is significantly
higher (but remarkably constant around $\sim 30$~\kms) in plage regions. Comparison
with non-thermal line broadening of the \Oline\
shows that it is smaller than 10~\kms\ and
that there is a general correspondence on
large spatial scales, 
but not on small spatial scales. This suggests that both 
increased microturbulence as well as so-called opacity broadening must play a
role in the large wing widths of \MgIIk. The opacity broadening comes
about because \MgIIk\ is an optically thick line, with the peak
separation of \ktwo\ strongly influenced by the chromospheric column mass.
There is a significant sensitivity of the wing
width to chromospheric temperatures (as well as microturbulence).

The combination of these diagnostics with our numerical experiments indicates that 
the chromospheric temperatures in plage are remarkably constant and
likely of the order of 6,000-6,500 K. This is compatible with the observed radiation
temperatures of \ktwo, and further confirmed by the fact that
the subordinate blend at 279.88 nm is rarely found to be in emission,
something that would be expected for higher temperatures.

The nature of the significant microturbulence in plage remains
unknown, but clearly the presence of strong shocks \citep[driving dynamic
fibrils and affecting line broadening;][]{2015ApJ...799L..12D}, torsional motions \citep{2014Sci...346D.315D} in and around plage, and Alfv\'en wave turbulence
\citep{2011ApJ...736....3V}
are candidates.

Currently ongoing numerical experiments suggest that the observed \MgIIkprofiles\ in
plage can only be explained by the combination of a TR at
high column mass (i.e., hot, dense corona above) as well as strong
chromospheric heating. Without the latter, densities at elevated
chromospheric temperatures would be lower, thus leading to 
narrower profiles
and a lack of filled-in \MgIIkprofiles. 

More work is required to better understand the coupling between
chromospheric and coronal conditions in plage, as well as the apparent
difficulty of reconciling the plage atmosphere considered here with
observations in \Cair\ and \Halpha, which may necessitate models with
multiple atmospheric components, perhaps similar to those found from
modeling of \CaIIHK\
\citep{1991A&A...250..220S} 
and \Cair\
\citep{2013ApJ...764L..11D}. 
\begin{acknowledgements}
This research has received funding from the European Research
Council 
(FP7/2007-2013)/ERC Grant agreement No.\ 291058, the Research Council of Norway,
the Swedish Knut and Alice
Wallenberg foundation, and NASA contract NNG09FA40C (IRIS), and has benefited from discussions at the International Space Science Institute (ISSI).
Thanks to Luc Rouppe van der Voort for support with
the SST observations. IRIS is a NASA small explorer developed and operated by LMSAL with mission operations executed at NASA Ames and major contributions to downlink communications funded by ESA and the Norwegian Space Centre.
The
Swedish 1-m Solar Telescope is operated by the Institute for Solar
Physics of Stockholm University in the Spanish
Observatorio del Roque de los Muchachos of the Instituto de
Astrof\'{\i}sica de Canarias.
\end{acknowledgements}

\bibliographystyle{apj}

\end{document}